\documentclass[twocolumn,showpacs,amsmath,prl,aps,amssymb,superscriptaddress]{revtex4-1} 
\usepackage[T1]{fontenc}
\usepackage[latin9]{inputenc}
\usepackage{times}
\usepackage{color} 
\usepackage{xspace}
\usepackage{amssymb,amsmath}
\usepackage{amsbsy}
\usepackage[pdftex]{graphicx}
\usepackage{bm}
\usepackage{float}

\usepackage[unicode,breaklinks]{hyperref}
\hypersetup{
    unicode=true,
    plainpages=false, 
    colorlinks=true,
    linkcolor=blue,
    citecolor=blue,
    filecolor=black,
    urlcolor=blue
}
\usepackage{url}
\usepackage{verbatim}

\newcommand{\br}{\mathbf{r}}

\begin{document}

\title{Crystallisation of a dilute atomic dipolar condensate}
\author{R.~N.~Bisset}  
\affiliation{Dodd-Walls Centre for Photonic and Quantum Technologies, Department of Physics, University of Otago, Dunedin, New Zealand}
\author{P.~B.~Blakie}  
\affiliation{Dodd-Walls Centre for Photonic and Quantum Technologies, Department of Physics, University of Otago, Dunedin, New Zealand}

\begin{abstract} 
We present a theory that explains the  experimentally observed crystallisation of a dilute dysprosium condensate into a  lattice of droplets. The key ingredient of our theory is a conservative three-body interaction which stabilises the droplets against collapse to  high density spikes. Our theory reproduces the experimental observations, and provides insight into the manybody properties of this new phase of matter. Notably, we show that it is unlikely that a supersolid was obtained in experiments, however our results suggest a strategy to realize this phase.  
\end{abstract}
 

\maketitle

\begin{figure}[htbp]
   \centering
   \includegraphics[width=3.3in]{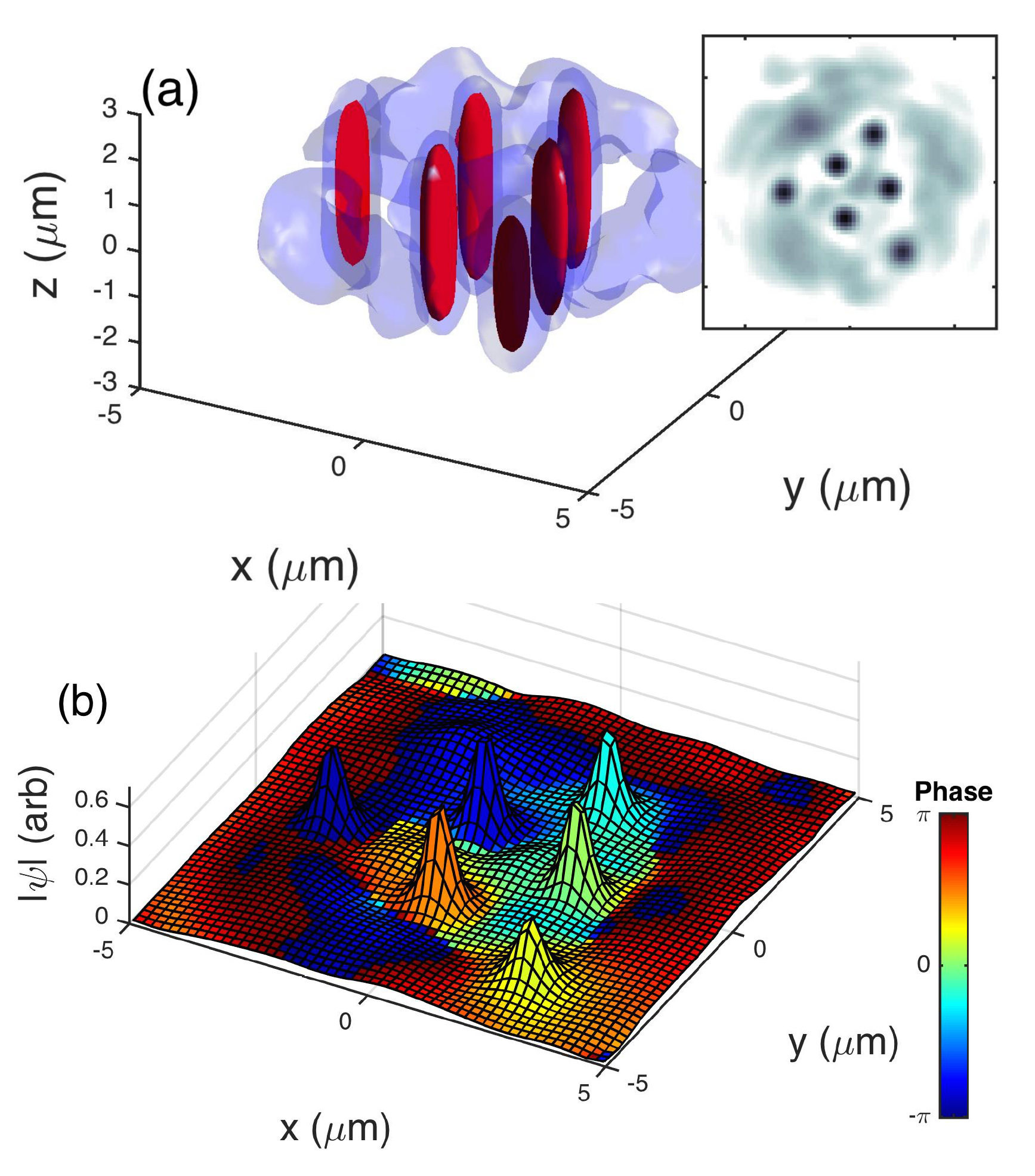} 
   \caption{(a) Droplet crystal obtained in our simulations at $t=15\,m$s after the quench is started. Red surface indicates high density isosurface at $n=2\times10^{20}\,$m$^{-3}$ and the blue low density isosurface is $n=0.2\times10^{20}\,$m$^{-3}$. The inset is a column density made by integrating the density along the $z$ axis.
    (b) Condensate phase in the $z=0$ plane.  Simulation for  $N_{\mathrm{cond}}=15\times10^3$, $T=20\,n$K, and $(\kappa_r,\kappa_i)=(5.87\times10^{-39},7.8\times10^{-42})$m$^6/$s,   with  other parameters as discussed in the text.}
   \label{fig1}
\end{figure}

Recent experiments with ultra-cold dysprosium \cite{Kadau2015a} have observed the crystallisation of a superfluid into a regular array of droplets. This result is surprising because it realizes a droplet crystal, which has been the subject of considerable theoretical work, yet avoids much of the complex interaction engineering in existing proposals, e.g.~using ensembles of cold Rydberg atoms \cite{Henkel2010a,Cinti2010a} or  polar molecules \cite{Lu2015a}. However, the standard theoretical description of dysprosium condensates does not predict a stable droplet phase, suggesting the role of additional physics. In this paper we propose that a conservative three-body interaction (e.g.~see \cite{Kohler2002a}) is the essential new physics. Augmenting the standard theory with this term we show that a stable droplet crystal forms (e.g.~see Fig.~\ref{fig1}) and we are able to explain the main observations made in the experiment \cite{Kadau2015a}. We note that this type of interaction has been found to play a dominant role in recent experiments with $^{85}$Rb \cite{Everitt2015a}.

Dysprosium  has a large magnetic moment giving rise to a significant long-ranged dipole-dipole interaction (DDI) between the atoms. In the experiment it is necessary to enhance (via Feshbach resonance) the short range repulsive interaction to be comparable in strength to the DDI in order to produce the initial (unstructured) condensate. The crystallisation is then initiated by suddenly reducing the value of the short range interaction, allowing the condensate to evolve with a dominant DDI. Detailed observations of the crystallisation dynamics are revealed by the use of high-resolution \textit{in situ} imaging of the system. Key experimental observations include: (i) the droplets form into an approximately triangular lattice with a lattice constants in the range $2$ to $3\,\mu$m and persists for long times ($\gg100$ ms, although with some droplet dynamics); (ii) the lattice formation time is  $\sim$7 ms after the interaction quench; (iii) the number of droplets formed is stochastic, but on average (over the range studied) the average droplet number increases linearly with condensate number. 

Because the  system we consider is dilute its dynamics should be well-described by the meanfield Gross-Pitaevskii theory. This theory includes short-range (contact) and long-range (dipole-dipole) two-body interactions \cite{Lahaye2009a}, and has been successful at describing a range of equilibrium and dynamic phenomena such as  the  parameter regions where the condensate is mechanically stable \cite{Koch2008a,Wilson2009a} and collapse dynamics \cite{Lahaye2009a,Muller2011a}. However, this theory applied to the experimental scenario outlined above fails to describe the observed dynamics: meanfield theory predicts that the system is unstable to forming sharp infinite-density spikes \cite{Linscott2014a}, in which regime the meanfield description is invalid. Indeed, such mechanical collapse scenarios are usually accompanied by explosive dynamics, rapid atom loss and heating (e.g.~see \cite{Sackett1998a,Donley2001a,Lahaye2009a}). Hence the observation of regular, stable and long-lived droplets in the dilute regime indicates the presence of additional physics not included in the standard theory.

Three-body recombination is an important loss mechanism in experiments which occurs when three ultra-cold atoms collide to form a di-atomic molecule and an atom that are both  lost from the atom trap.   The measurement of the three-body recombination rate has been used to reveal quantum statistical and manybody effects (e.g.~see \cite{Burt1997a,Tolra2004a}), and to locate Feshbach resonances (e.g.~see \cite{Maier2015a}). There is also a conservative three-body interaction between the particles, which does not lead to loss \cite{Kohler2002a}, and is expected to be large if there is an Efimov state near the collision threshold \cite{Braaten2002a,Bulgac2002a}. While we are not aware of any quantitative predictions for three-body interactions in  $^{164}$Dy, this atom is known to have complex collisional properties, including a large number of Feshbach resonances \cite{Maier2015a}.
Since ultra-cold atomic systems are so dilute the role of such three-body interactions is typically much smaller than the two-body interactions. However, in scenarios where the system becomes mechanically unstable due to attractive two-body interactions the density can increase significantly and three-body terms can be important \cite{Bulgac2002a}. Indeed, a recent experiment with $^{85}$Rb \cite{Everitt2015a} has measured a conservative three-body interaction that is 2 to 3 orders of magnitude larger than the three-body recombination rate.  
Here we show that by including a three-body interaction of comparable size we are able to quantitatively describe the crystallisation observed in the dysprosium experiments.
  
To perform simulations we take the system evolution to be described by the Gross-Pitaevskii equation  \begin{align}
i\hbar\frac{\partial \psi}{\partial t}\!=\!&\left[H_{\mathrm{sp}}\!+\!\int\!d\br'\,U(\br\!-\!\br',t)|\psi(\br',t)|^2+\frac{\kappa_3}{2}|\psi|^4\right]\psi,\label{GPE}
\end{align}
where $H_{\mathrm{sp}}={-\hbar^2\nabla^2}/{2m}+V_{\mathrm{trap}}$, and
\begin{align}
U(\mathbf{r},t)=\frac{4\pi\hbar^2a(t)}{m}\delta(\br)+\frac{\mu_0\mu^2}{4\pi}\frac{1-3\cos^2\theta}{r^3}
\end{align}
describes the two-body contact and dipolar interactions for dipoles polarized along $z$, with $\theta$ being the angle between $\br$ and the $z$-axis. Here $m$ is the atomic mass and $\mu=9.93\mu_{\mathrm{B}}$ is the magnetic moment of a Dy atom, with $\mu_{\mathrm{B}}$ the Bohr magneton. The two-body contact interaction, parameterized by the $s$-wave scattering length  $a(t)$, is time-dependent as it is changed using a magnetic Feshbach resonance. The last term in (\ref{GPE}) describes short-ranged three-body interactions. The coefficient is complex $\kappa_3=\kappa_r-i\kappa_i$, with  $\kappa_r$ characterizing the strength of the conservative component and $\kappa_i$ quantifying the three-body recombination loss rate.

We perform simulations in the regime reported in Ref.~\cite{Kadau2015a} and consider  condensates up to $N_{\mathrm{cond}}=20\times10^3$  atoms prepared in a harmonic trap ($V_{\mathrm{trap}}$) with frequencies $(\nu_x,\nu_y,\nu_z)=(45,45,133)$ Hz and with the dipoles polarized along the $z$ axis. 
The condensate is initially prepared with a scattering length of $a_i=130 a_0$, where $a_0$ is the Bohr radius. This value is obtained using a Feshbach resonance in the experiment, and ensures that a stable (unstructured) condensate is produced.

We take the initial condensate $\psi_0(\br)$ to be the stationary solution of Eq.~(\ref{GPE}) normalized to $N_{\mathrm{cond}}$ with $\kappa_3=0$, which we solve for using a Newton-Krylov scheme \cite{Ronen2006a,Martin2012a}. We note that the effect of the $\kappa_r$ values we use for dynamics is negligible in the initial state (with peak density of $\sim0.91\times10^{20}$m$^{-3}$ for $N_{\mathrm{cond}}=15\times10^3$), making less than a 1\% change in the ground state energy. Thus taking $\kappa_r=0$ for the initial state preparation is a good approximation.
To the condensate we add initial state fluctuations to account for quantum and thermal fluctuations in the system. Such fluctuations usually play an important role in seeding unstable dynamics, and are added as
\begin{equation}
\psi(\br,0)=\psi_0(\br)+{\sum_{n}}'\alpha_n\phi_n(\br),\label{noise}
\end{equation}
where $\epsilon_n\phi_n=H_{\mathrm{sp}}\phi_n$ are the single particle eigenstates, $\alpha_n$ is a complex gaussian random variable with $\langle |\alpha_n|^2\rangle=(e^{\epsilon_n/k_BT}-1)^{-1}+\frac{1}{2}$, and the sum in (\ref{noise}) is restricted to modes with $\epsilon_n\le 2k_BT$. This choice of fluctuations is according to the truncated Wigner prescription (see \cite{cfieldRev2008}) for a system at temperature $T$. The main results we present are for $T= 20\,n$K, adding approximately 400 atoms to the system, consistent with the experimental conditions of a ``quasi-pure condensate" (c.f.~the ideal condensation temperature of $T_c=72\,n$K for $N=15\times10^3$). 
 For dynamics we evolve the system according to the GPE (\ref{GPE}) discretised on a three-dimensional grid in a cubic box of dimension $23.4\,\mu$m, propagated in time using a 4th order Runge-Kutta integration method. The number of grid points is varied to check accuracy of results.
The kinetic energy term is evaluated in Fourier space for spectral accuracy, and the DDI term is evaluated using Fast Fourier transforms to action the convolution, with a spherically cut-off dipole kernel used in $k$-space to minimise boundary effects \cite{Ronen2006a}. 
The three-body interaction only plays an important role in the dynamics when the density gets high and we choose to use a constant value of $\kappa_3$ throughout each simulation.
 
\begin{figure}[htbp]
   \centering
   \includegraphics[width=3.0in]{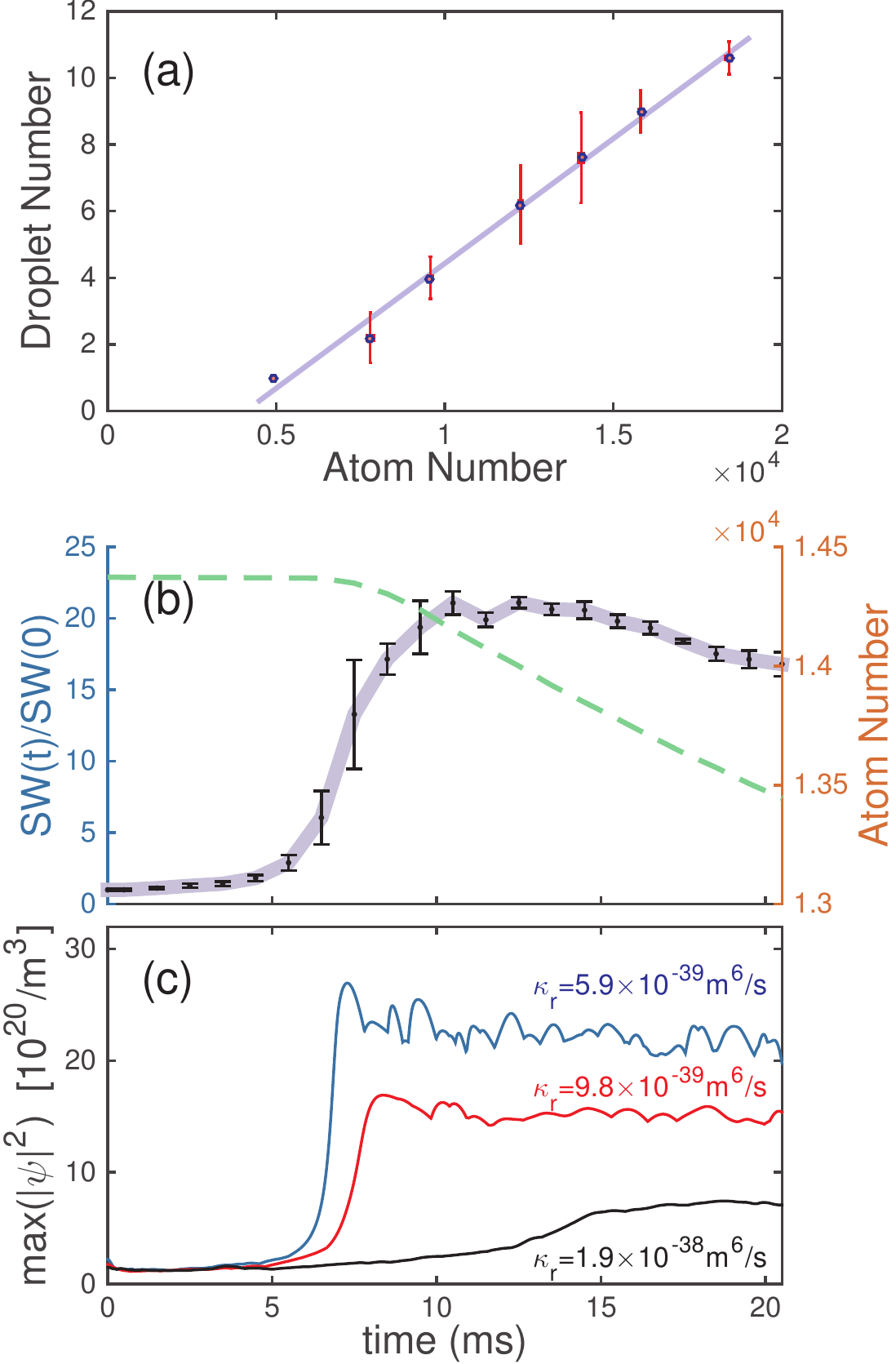} 
   \caption{Droplet formation. (a) Number of droplets versus total atom number averaged over 5 simulations for each case. Line is best-fit with slope $7.50\times10^{-4}$. (b) Relative spectral weight versus time. (c) Peak density for simulations with different $\kappa_r$ values. In all cases $T=20\,n$K and $\kappa_i=7.8\times10^{-42}$m$^{6}/$s.  In (a) and (b) $\kappa_r=5.87\times10^{-39}$m$^{6}/$s. In (b) and (c)  $N_{\mathrm{cond}}=14\times10^3$.  }
   \label{figdroplets}
\end{figure}

The $s$-wave scattering length is linearly ramped from $a_i$ to $a_f$ over 0.5 ms, and then held constant for the remainder of the simulation.
As in experiment we consider a quench to $a_f$, close to the background value $a_{\mathrm{bg}}$, which initiates the crystallisation.
There  remains appreciable uncertainty in the value of $a_{\mathrm{bg}}$ with the current experimental estimates being $92(8)\,a_0$ \cite{Tang2015a,Maier2015a}. Having explored a range of simulation parameters we find  dynamics similar to experiment  for $a_f\approx 82.6\,a_0$ \footnote{For $a_f\gtrsim92\,a_0$ we do not find any droplets forming within $15\,m$s, and on longer timescales much few droplets form than observed in experiment.}.
At this value the condensate is susceptible to the growth of unstable modes \cite{Santos2003a,Ronen2007a,Blakie2012a}; these lead to the development of high density regions  near the trap centre, and then across the condensate, driven by the attractive component of the DDI. On a timescale of 5--15 ms (depending on parameters) several high density droplets form and the role of $\kappa_3$ becomes crucially important [e.g.~see Fig.~\ref{figdroplets}(c)]. 

We automate droplet detection in our simulations by identifying local column-density maxima  (with densities  exceeding 1.2 $\times$ the peak density of the initial condensate). We identify the region about this point where the density decays to define the droplet. We note (e.g.~see Fig.~\ref{fig1}) that once the droplets fully form they deplete the atomic density at their boundaries and are thus unambiguously  identified.

In Fig.~\ref{figdroplets}(a) we show results for the number of droplets that are identified at 20 ms after the quench begins as a function of the atom number. Results are from 5 calculations at each atom number, and with an initial temperature of $T=20\,n$K and we take $(\kappa_r,\kappa_i)=(5.87\times10^{-39},7.8\times10^{-42})$m$^6/$s.   
Because of the different initial noise the number of droplets that form vary from run to run. Our results are similar to  the experimental measurements \cite{Kadau2015a}, except that the experimental measurements tend to find more droplets forming for $5\times10^3$ atoms, which may indicate that either $a_f$ is  lower than we use, or our criteria for identifying droplets are different.

Following the experiment we evaluate a spectral weight  to quantify the spatial structure in the \textit{in situ} column density images on $\mu$m length scales post-quench. This is done in terms of the function $\mathrm{SW}(t)=\sum_k'S_t(k)$, where $S_t(k)$ is the radially averaged Fourier transform of the  column density (along $z$) at time $t$,  with the sum  taken over  the range $k\in[1.5, 5]\,\mu\mathrm{m}^{-1}$. The relative spectral weight, $\mathrm{SW}(t)/\mathrm{SW}(0)$, is a measure of the  increase in the density modulation post-quench relative to the initial condensate. We find that the relative spectral weight increases to much larger values than those measured in experiment, because we do not account for the finite resolution effects  of the imaging system. However, we obtain good quantitative agreement with experiment for the timescales over which the spectral weight grows and subsequently decays [see Fig.~\ref{figdroplets}(b)]. This decay arises from  three-body recombination and a collective breathing mode of the droplet crystal that is excited during formation. The rate of atom loss is more rapid in the droplet phase than the original condensate because of the significantly higher density.

We have conducted simulations  exploring a wide range of parameters, and the values of $a_{f}$ and $\kappa_3$ used above were determined to give a reasonable fit to the experimental results.  
More  information from experiments is necessary to completely determine these parameters. Detailed results for the formation dynamics [see Fig.~\ref{figdroplets}(c)] shows that the droplet formation time is sensitive to the value of $\kappa_r$. Indeed, the largest value used in Fig.~\ref{figdroplets}(c) leads to a formation time that is at least twice as long as that seen in experiment (as well as leading to too few droplets $\sim3$). 
 We have also explored the effect of changing temperature and find only a slight change in formation time for temperatures up to $T\sim40\,n$K.  
 
 We also find that the droplet size and peak density are sensitive to $\kappa_r$.   
This can be simply understood because the droplet forms from the competition between the attractive two-body and the repulsive three-body interactions. The density at which this balance is achieved scales inversely with $\kappa_r$. We find that the droplets that form near the center of the trap have very similar properties, including the atom number in each droplet,  peak density, transverse [$w_{x,y}$] and axial [$w_z$] widths, where 
\begin{equation}
w_\nu^2\equiv \gamma N_D^{-1} {\int_{\mathrm{D}}d\mathbf{r}\,(r_\nu-r_\nu^{\mathrm{c}})^2|\psi(\mathbf{r})|^2},\quad\nu=\{x,y,z\} , \label{Dropletwidths}
 \end{equation}
with, the integration restricted to the region containing the droplet, $\mathbf{r}^{\mathrm{c}}$ the center of the droplet, and $N_{\mathrm{D}}$ the number of atoms in the droplet. The factor $\gamma=8\ln(2)$  is chosen to calibrate this width measure to be the full width at half maximum for gaussian shaped droplets.
In Table \ref{tab:droplets} we show how the properties of the droplets change with the three-body interaction. Notably, the droplets get wider and less dense, but hold more atoms, as $\kappa_r$ increases.

\begin{table}[htbp]
   \centering 
   \begin{tabular}{ p{1.9cm}|  p{0.25cm} p{1.5cm} p{1.5cm} p{1.5cm} p{1.5cm}p{1.5cm}} 
    \hline\hline  
    
    
      $\kappa_r$	&     &  $N_{\mathrm{D}}$  &       $n_{\mathrm{peak}}$   &  $w_{x,y}$  &     $w_z$ \\  
      $10^{-39}\,$m$^{6}/$s& & $10^3$  & $10^{20}\,\mathrm{m}^{-3}$ & $10^{-6}\,\mathrm{m}$ & $10^{-6}\,\mathrm{m}$ \\
      \hline 
     3.91	&	&  1.2(1)	&  28(2)   &  0.26(1) 	&  2.5(2)      \\ 
      5.87	  &      &  1.5(1)	& 20(2)    &  0.33(2) 	&  2.7(1)  \\ 
    7.83 	&	&  1.8(2)	& 18(1)    &  0.36(2) &  2.9(2)        \\ 
     9.78	&	&  2.0(2)	& 14(1)   	&  0.42(3)	& 2.9(2)      \\ 
      11.7	&	&  2.4(4)	&  13(2) 	&  0.46(2)	& 3.1(2)      \\  
       19.6	&	&  3.4(8)	&  7(1) 	&  0.60(1)	& 3.3(2)      \\  
    \hline
   \end{tabular}
   \caption{Droplet properties as $\kappa_r$ varies for $N_{\mathrm{cond}}=15\times10^3$, $a_f=82.6\,a_0$, with $T=20\,n$K and $\kappa_i=7.8\times10^{-42}$m$^6/$s. Evaluated for droplets located within $2.5\,\mu$m of the trap center (in the $xy$-plane) using simulation results for $t\ge18\,m$s. The error indicates the standard deviation across measured droplets.}
   \label{tab:droplets}
\end{table}

\begin{figure}[htbp]
   \centering
   \includegraphics[width=3.4in]{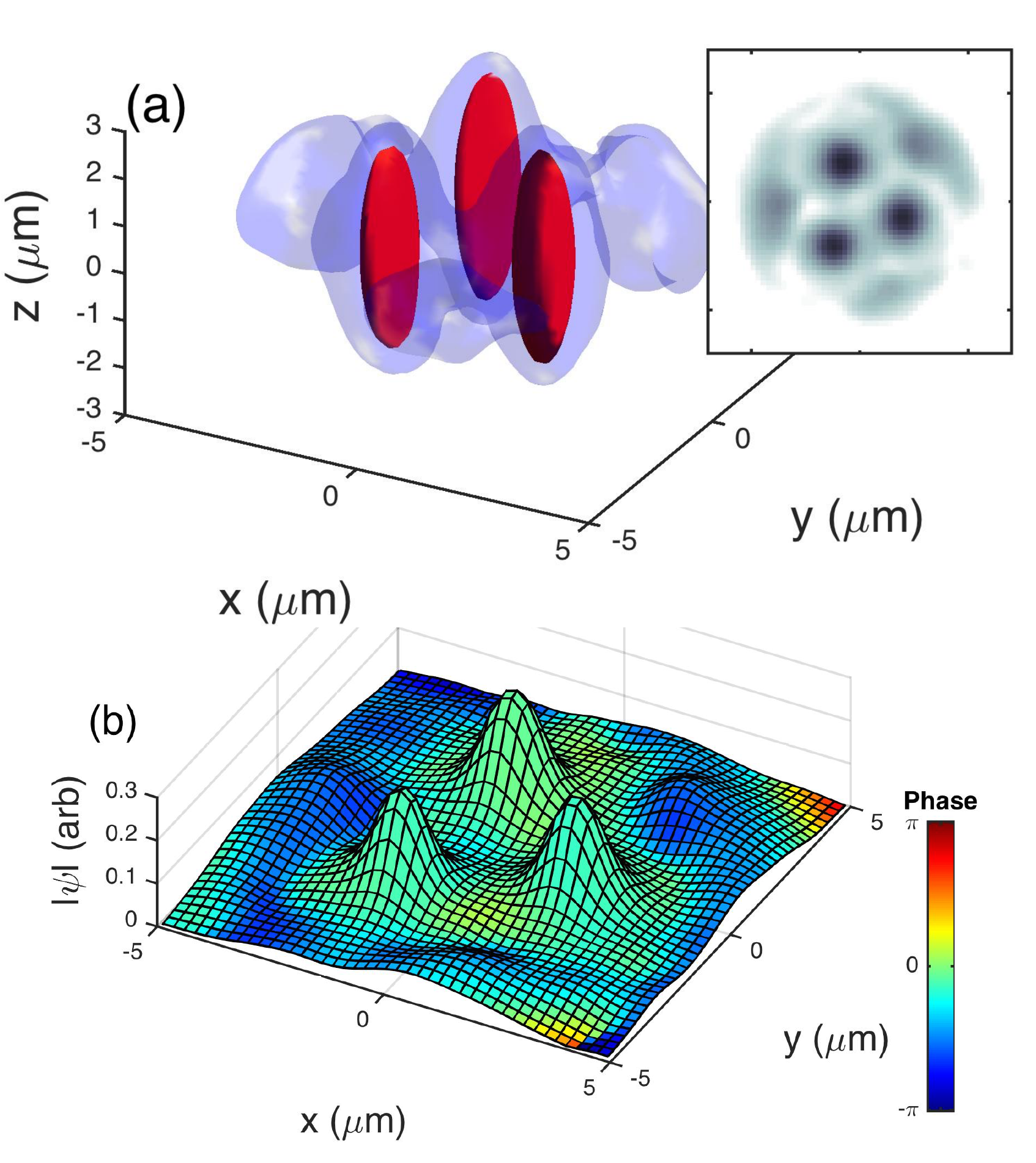} 
   \caption{(a) Droplet crystal obtained in our simulations at $15\,m$s after the quench is started. Red surface indicates high density isosurface at $n=2\times10^{20}$m$^{-3}$ and the blue low density isosurface is $n=0.2\times10^{20}$m$^{-3}$. The inset is a column density made by integrating the density along the $z$ axis.
    (b) The condensate phase in the $z=0$ plane. Simulation for $15\times10^3$atoms, $\kappa_r=1.96\times 10^{-38}$ m$^6$/s.}
   \label{figSS}
\end{figure}

An important area of interest is whether this droplet crystal maintains phase coherence, and hence could be a supersolid \cite{Kim2004a,Kim2012a,Boninsegni2012a}. This question was not able to be explored in the experiment. 
 We have examined this by analysing the mean phase of each droplet (the phase varies minimally within each droplet). We find that the first droplets formed have similar phases, which they inherit  from the condensate, but in general these quickly become independent. For example, the state shown in Fig.~\ref{fig1} is  found to develop independent phases by $\sim15\,m$s after the quench [see Fig.~\ref{fig1}(b)]. This occurs due to heating during the droplet formation, e.g.~we observe vortex anti-vortex pairs created between droplets and the creation of additional phonon excitations.
  Thus, we predict it is unlikely that the state produced in experiment is a supersolid. However, using a larger value of $\kappa_r=1.96\times 10^{-38}\,$m$^{6}/$s  we have observed the formation of a small droplet crystal in which the phase coherence persists for at least $400\,m$s [see Fig.~\ref{figSS}(b)]. 
In this case the droplets are approximately twice as wide  [see Table \ref{tab:droplets}] compared to the case in Fig.~\ref{fig1}(b), only 3 drops form and the formation time is much slower [$\sim15\,m$s, see Figs,~\ref{figdroplets}(c) and \ref{figSS}(a)], and causes less heating. The reduced heating and enhanced tunnelling between the larger droplets allows this system to behave as a supersolid. 

In conclusion we have identified the key physical mechanism behind the recent observation of a droplet crystal in a dilute gas of dysprosium. We have investigated properties of the droplets that form and important factors in maintaining phase coherence of the crystal in order to produce a supersolid. Our results show that the crystal produced in experiments was likely not a supersolid because of heating during its rapid formation.
 In future work we will explore experimentally practical adjustments (changing two-body interactions, trap geometry etc.) which will enable larger crystals to form in the regime where phase coherence is maintained between the droplets.
 
We gratefully acknowledge support from the Marsden Fund of the Royal Society of New Zealand.  We also note that in the final stages of preparing this manuscript for submission a pre-print appeared that also investigates three-body interactions in relation to the formation of a droplet phase in a dipolar condensate \cite{Xi2015a}.


\begin{thebibliography}{30}%
\makeatletter
\providecommand \@ifxundefined [1]{%
 \@ifx{#1\undefined}
}%
\providecommand \@ifnum [1]{%
 \ifnum #1\expandafter \@firstoftwo
 \else \expandafter \@secondoftwo
 \fi
}%
\providecommand \@ifx [1]{%
 \ifx #1\expandafter \@firstoftwo
 \else \expandafter \@secondoftwo
 \fi
}%
\providecommand \natexlab [1]{#1}%
\providecommand \enquote  [1]{``#1''}%
\providecommand \bibnamefont  [1]{#1}%
\providecommand \bibfnamefont [1]{#1}%
\providecommand \citenamefont [1]{#1}%
\providecommand \href@noop [0]{\@secondoftwo}%
\providecommand \href [0]{\begingroup \@sanitize@url \@href}%
\providecommand \@href[1]{\@@startlink{#1}\@@href}%
\providecommand \@@href[1]{\endgroup#1\@@endlink}%
\providecommand \@sanitize@url [0]{\catcode `\\12\catcode `\$12\catcode
  `\&12\catcode `\#12\catcode `\^12\catcode `\_12\catcode `\%12\relax}%
\providecommand \@@startlink[1]{}%
\providecommand \@@endlink[0]{}%
\providecommand \url  [0]{\begingroup\@sanitize@url \@url }%
\providecommand \@url [1]{\endgroup\@href {#1}{\urlprefix }}%
\providecommand \urlprefix  [0]{URL }%
\providecommand \Eprint [0]{\href }%
\providecommand \doibase [0]{http://dx.doi.org/}%
\providecommand \selectlanguage [0]{\@gobble}%
\providecommand \bibinfo  [0]{\@secondoftwo}%
\providecommand \bibfield  [0]{\@secondoftwo}%
\providecommand \translation [1]{[#1]}%
\providecommand \BibitemOpen [0]{}%
\providecommand \bibitemStop [0]{}%
\providecommand \bibitemNoStop [0]{.\EOS\space}%
\providecommand \EOS [0]{\spacefactor3000\relax}%
\providecommand \BibitemShut  [1]{\csname bibitem#1\endcsname}%
\let\auto@bib@innerbib\@empty
\bibitem [{\citenamefont {{Kadau}}\ \emph {et~al.}(2015)\citenamefont
  {{Kadau}}, \citenamefont {{Schmitt}}, \citenamefont {{Wenzel}}, \citenamefont
  {{Wink}}, \citenamefont {{Maier}}, \citenamefont {{Ferrier-Barbut}},\ and\
  \citenamefont {{Pfau}}}]{Kadau2015a}%
  \BibitemOpen
  \bibfield  {author} {\bibinfo {author} {\bibfnamefont {H.}~\bibnamefont
  {{Kadau}}}, \bibinfo {author} {\bibfnamefont {M.}~\bibnamefont {{Schmitt}}},
  \bibinfo {author} {\bibfnamefont {M.}~\bibnamefont {{Wenzel}}}, \bibinfo
  {author} {\bibfnamefont {C.}~\bibnamefont {{Wink}}}, \bibinfo {author}
  {\bibfnamefont {T.}~\bibnamefont {{Maier}}}, \bibinfo {author} {\bibfnamefont
  {I.}~\bibnamefont {{Ferrier-Barbut}}}, \ and\ \bibinfo {author}
  {\bibfnamefont {T.}~\bibnamefont {{Pfau}}},\ }\href@noop {} {\bibfield
  {journal} {\bibinfo  {journal} {ArXiv e-prints}\ } (\bibinfo {year}
  {2015})},\ \Eprint {http://arxiv.org/abs/1508.05007} {arXiv:1508.05007
  [cond-mat.quant-gas]} \BibitemShut {NoStop}%
\bibitem [{\citenamefont {Henkel}\ \emph {et~al.}(2010)\citenamefont {Henkel},
  \citenamefont {Nath},\ and\ \citenamefont {Pohl}}]{Henkel2010a}%
  \BibitemOpen
  \bibfield  {author} {\bibinfo {author} {\bibfnamefont {N.}~\bibnamefont
  {Henkel}}, \bibinfo {author} {\bibfnamefont {R.}~\bibnamefont {Nath}}, \ and\
  \bibinfo {author} {\bibfnamefont {T.}~\bibnamefont {Pohl}},\ }\href {\doibase
  10.1103/PhysRevLett.104.195302} {\bibfield  {journal} {\bibinfo  {journal}
  {Phys. Rev. Lett.}\ }\textbf {\bibinfo {volume} {104}},\ \bibinfo {pages}
  {195302} (\bibinfo {year} {2010})}\BibitemShut {NoStop}%
\bibitem [{\citenamefont {Cinti}\ \emph {et~al.}(2010)\citenamefont {Cinti},
  \citenamefont {Jain}, \citenamefont {Boninsegni}, \citenamefont {Micheli},
  \citenamefont {Zoller},\ and\ \citenamefont {Pupillo}}]{Cinti2010a}%
  \BibitemOpen
  \bibfield  {author} {\bibinfo {author} {\bibfnamefont {F.}~\bibnamefont
  {Cinti}}, \bibinfo {author} {\bibfnamefont {P.}~\bibnamefont {Jain}},
  \bibinfo {author} {\bibfnamefont {M.}~\bibnamefont {Boninsegni}}, \bibinfo
  {author} {\bibfnamefont {A.}~\bibnamefont {Micheli}}, \bibinfo {author}
  {\bibfnamefont {P.}~\bibnamefont {Zoller}}, \ and\ \bibinfo {author}
  {\bibfnamefont {G.}~\bibnamefont {Pupillo}},\ }\href {\doibase
  10.1103/PhysRevLett.105.135301} {\bibfield  {journal} {\bibinfo  {journal}
  {Phys. Rev. Lett.}\ }\textbf {\bibinfo {volume} {105}},\ \bibinfo {pages}
  {135301} (\bibinfo {year} {2010})}\BibitemShut {NoStop}%
\bibitem [{\citenamefont {Lu}\ \emph {et~al.}(2015)\citenamefont {Lu},
  \citenamefont {Li}, \citenamefont {Petrov},\ and\ \citenamefont
  {Shlyapnikov}}]{Lu2015a}%
  \BibitemOpen
  \bibfield  {author} {\bibinfo {author} {\bibfnamefont {Z.-K.}\ \bibnamefont
  {Lu}}, \bibinfo {author} {\bibfnamefont {Y.}~\bibnamefont {Li}}, \bibinfo
  {author} {\bibfnamefont {D.~S.}\ \bibnamefont {Petrov}}, \ and\ \bibinfo
  {author} {\bibfnamefont {G.~V.}\ \bibnamefont {Shlyapnikov}},\ }\href
  {\doibase 10.1103/PhysRevLett.115.075303} {\bibfield  {journal} {\bibinfo
  {journal} {Phys. Rev. Lett.}\ }\textbf {\bibinfo {volume} {115}},\ \bibinfo
  {pages} {075303} (\bibinfo {year} {2015})}\BibitemShut {NoStop}%
\bibitem [{\citenamefont {K\"ohler}(2002)}]{Kohler2002a}%
  \BibitemOpen
  \bibfield  {author} {\bibinfo {author} {\bibfnamefont {T.}~\bibnamefont
  {K\"ohler}},\ }\href {\doibase 10.1103/PhysRevLett.89.210404} {\bibfield
  {journal} {\bibinfo  {journal} {Phys. Rev. Lett.}\ }\textbf {\bibinfo
  {volume} {89}},\ \bibinfo {pages} {210404} (\bibinfo {year}
  {2002})}\BibitemShut {NoStop}%
\bibitem [{\citenamefont {{Everitt}}\ \emph {et~al.}(2015)\citenamefont
  {{Everitt}}, \citenamefont {{Sooriyabandara}}, \citenamefont {{McDonald}},
  \citenamefont {{Hardman}}, \citenamefont {{Quinlivan}}, \citenamefont
  {{Perumbil}}, \citenamefont {{Wigley}}, \citenamefont {{Debs}}, \citenamefont
  {{Close}}, \citenamefont {{Kuhn}},\ and\ \citenamefont
  {{Robins}}}]{Everitt2015a}%
  \BibitemOpen
  \bibfield  {author} {\bibinfo {author} {\bibfnamefont {P.~J.}\ \bibnamefont
  {{Everitt}}}, \bibinfo {author} {\bibfnamefont {M.~A.}\ \bibnamefont
  {{Sooriyabandara}}}, \bibinfo {author} {\bibfnamefont {G.~D.}\ \bibnamefont
  {{McDonald}}}, \bibinfo {author} {\bibfnamefont {K.~S.}\ \bibnamefont
  {{Hardman}}}, \bibinfo {author} {\bibfnamefont {C.}~\bibnamefont
  {{Quinlivan}}}, \bibinfo {author} {\bibfnamefont {M.}~\bibnamefont
  {{Perumbil}}}, \bibinfo {author} {\bibfnamefont {P.}~\bibnamefont
  {{Wigley}}}, \bibinfo {author} {\bibfnamefont {J.~E.}\ \bibnamefont
  {{Debs}}}, \bibinfo {author} {\bibfnamefont {J.~D.}\ \bibnamefont {{Close}}},
  \bibinfo {author} {\bibfnamefont {C.~C.~N.}\ \bibnamefont {{Kuhn}}}, \ and\
  \bibinfo {author} {\bibfnamefont {N.~P.}\ \bibnamefont {{Robins}}},\
  }\href@noop {} {\bibfield  {journal} {\bibinfo  {journal} {ArXiv e-prints}\ }
  (\bibinfo {year} {2015})},\ \Eprint {http://arxiv.org/abs/1509.06844}
  {arXiv:1509.06844 [cond-mat.quant-gas]} \BibitemShut {NoStop}%
\bibitem [{\citenamefont {Lahaye}\ \emph {et~al.}(2008)\citenamefont {Lahaye},
  \citenamefont {Metz}, \citenamefont {Fr\"{o}hlich}, \citenamefont {Koch},
  \citenamefont {Meister}, \citenamefont {Griesmaier}, \citenamefont {Pfau},
  \citenamefont {Saito}, \citenamefont {Kawaguchi},\ and\ \citenamefont
  {Ueda}}]{Lahaye2009a}%
  \BibitemOpen
  \bibfield  {author} {\bibinfo {author} {\bibfnamefont {T.}~\bibnamefont
  {Lahaye}}, \bibinfo {author} {\bibfnamefont {J.}~\bibnamefont {Metz}},
  \bibinfo {author} {\bibfnamefont {B.}~\bibnamefont {Fr\"{o}hlich}}, \bibinfo
  {author} {\bibfnamefont {T.}~\bibnamefont {Koch}}, \bibinfo {author}
  {\bibfnamefont {M.}~\bibnamefont {Meister}}, \bibinfo {author} {\bibfnamefont
  {A.}~\bibnamefont {Griesmaier}}, \bibinfo {author} {\bibfnamefont
  {T.}~\bibnamefont {Pfau}}, \bibinfo {author} {\bibfnamefont {H.}~\bibnamefont
  {Saito}}, \bibinfo {author} {\bibfnamefont {Y.}~\bibnamefont {Kawaguchi}}, \
  and\ \bibinfo {author} {\bibfnamefont {M.}~\bibnamefont {Ueda}},\ }\href
  {\doibase 10.1103/PhysRevLett.101.080401} {\bibfield  {journal} {\bibinfo
  {journal} {Phys. Rev. Lett.}\ }\textbf {\bibinfo {volume} {101}},\ \bibinfo
  {eid} {080401} (\bibinfo {year} {2008})}\BibitemShut {NoStop}%
\bibitem [{\citenamefont {Koch}\ \emph {et~al.}(2008)\citenamefont {Koch},
  \citenamefont {Lahaye}, \citenamefont {Metz}, \citenamefont {B.Froehlich},
  \citenamefont {Griesmaier},\ and\ \citenamefont {Pfau}}]{Koch2008a}%
  \BibitemOpen
  \bibfield  {author} {\bibinfo {author} {\bibfnamefont {T.}~\bibnamefont
  {Koch}}, \bibinfo {author} {\bibfnamefont {T.}~\bibnamefont {Lahaye}},
  \bibinfo {author} {\bibfnamefont {J.}~\bibnamefont {Metz}}, \bibinfo {author}
  {\bibnamefont {B.Froehlich}}, \bibinfo {author} {\bibfnamefont
  {A.}~\bibnamefont {Griesmaier}}, \ and\ \bibinfo {author} {\bibfnamefont
  {T.}~\bibnamefont {Pfau}},\ }\href@noop {} {\bibfield  {journal} {\bibinfo
  {journal} {Nat. Phys.}\ }\textbf {\bibinfo {volume} {4}},\ \bibinfo {pages}
  {218} (\bibinfo {year} {2008})}\BibitemShut {NoStop}%
\bibitem [{\citenamefont {Wilson}\ \emph {et~al.}(2009)\citenamefont {Wilson},
  \citenamefont {Ronen},\ and\ \citenamefont {Bohn}}]{Wilson2009a}%
  \BibitemOpen
  \bibfield  {author} {\bibinfo {author} {\bibfnamefont {R.~M.}\ \bibnamefont
  {Wilson}}, \bibinfo {author} {\bibfnamefont {S.}~\bibnamefont {Ronen}}, \
  and\ \bibinfo {author} {\bibfnamefont {J.~L.}\ \bibnamefont {Bohn}},\ }\href
  {\doibase 10.1103/PhysRevA.80.023614} {\bibfield  {journal} {\bibinfo
  {journal} {Phys. Rev. A}\ }\textbf {\bibinfo {volume} {80}},\ \bibinfo
  {pages} {023614} (\bibinfo {year} {2009})}\BibitemShut {NoStop}%
\bibitem [{\citenamefont {M\"uller}\ \emph {et~al.}(2011)\citenamefont
  {M\"uller}, \citenamefont {Billy}, \citenamefont {Henn}, \citenamefont
  {Kadau}, \citenamefont {Griesmaier}, \citenamefont {Jona-Lasinio},
  \citenamefont {Santos},\ and\ \citenamefont {Pfau}}]{Muller2011a}%
  \BibitemOpen
  \bibfield  {author} {\bibinfo {author} {\bibfnamefont {S.}~\bibnamefont
  {M\"uller}}, \bibinfo {author} {\bibfnamefont {J.}~\bibnamefont {Billy}},
  \bibinfo {author} {\bibfnamefont {E.~A.~L.}\ \bibnamefont {Henn}}, \bibinfo
  {author} {\bibfnamefont {H.}~\bibnamefont {Kadau}}, \bibinfo {author}
  {\bibfnamefont {A.}~\bibnamefont {Griesmaier}}, \bibinfo {author}
  {\bibfnamefont {M.}~\bibnamefont {Jona-Lasinio}}, \bibinfo {author}
  {\bibfnamefont {L.}~\bibnamefont {Santos}}, \ and\ \bibinfo {author}
  {\bibfnamefont {T.}~\bibnamefont {Pfau}},\ }\href {\doibase
  10.1103/PhysRevA.84.053601} {\bibfield  {journal} {\bibinfo  {journal} {Phys.
  Rev. A}\ }\textbf {\bibinfo {volume} {84}},\ \bibinfo {pages} {053601}
  (\bibinfo {year} {2011})}\BibitemShut {NoStop}%
\bibitem [{\citenamefont {Linscott}\ and\ \citenamefont
  {Blakie}(2014)}]{Linscott2014a}%
  \BibitemOpen
  \bibfield  {author} {\bibinfo {author} {\bibfnamefont {E.~B.}\ \bibnamefont
  {Linscott}}\ and\ \bibinfo {author} {\bibfnamefont {P.~B.}\ \bibnamefont
  {Blakie}},\ }\href {\doibase 10.1103/PhysRevA.90.053605} {\bibfield
  {journal} {\bibinfo  {journal} {Phys. Rev. A}\ }\textbf {\bibinfo {volume}
  {90}},\ \bibinfo {pages} {053605} (\bibinfo {year} {2014})}\BibitemShut
  {NoStop}%
\bibitem [{\citenamefont {Sackett}\ \emph {et~al.}(1998)\citenamefont
  {Sackett}, \citenamefont {Stoof},\ and\ \citenamefont
  {Hulet}}]{Sackett1998a}%
  \BibitemOpen
  \bibfield  {author} {\bibinfo {author} {\bibfnamefont {C.~A.}\ \bibnamefont
  {Sackett}}, \bibinfo {author} {\bibfnamefont {H.~T.~C.}\ \bibnamefont
  {Stoof}}, \ and\ \bibinfo {author} {\bibfnamefont {R.~G.}\ \bibnamefont
  {Hulet}},\ }\href {\doibase 10.1103/PhysRevLett.80.2031} {\bibfield
  {journal} {\bibinfo  {journal} {Phys. Rev. Lett.}\ }\textbf {\bibinfo
  {volume} {80}},\ \bibinfo {pages} {2031} (\bibinfo {year}
  {1998})}\BibitemShut {NoStop}%
\bibitem [{\citenamefont {Donley}\ \emph {et~al.}(2001)\citenamefont {Donley},
  \citenamefont {Claussen}, \citenamefont {Cornish}, \citenamefont {Roberts},
  \citenamefont {Cornell},\ and\ \citenamefont {Wieman}}]{Donley2001a}%
  \BibitemOpen
  \bibfield  {author} {\bibinfo {author} {\bibfnamefont {E.~A.}\ \bibnamefont
  {Donley}}, \bibinfo {author} {\bibfnamefont {N.~R.}\ \bibnamefont
  {Claussen}}, \bibinfo {author} {\bibfnamefont {S.~L.}\ \bibnamefont
  {Cornish}}, \bibinfo {author} {\bibfnamefont {J.~L.}\ \bibnamefont
  {Roberts}}, \bibinfo {author} {\bibfnamefont {E.~A.}\ \bibnamefont
  {Cornell}}, \ and\ \bibinfo {author} {\bibfnamefont {C.~E.}\ \bibnamefont
  {Wieman}},\ }\href@noop {} {\bibfield  {journal} {\bibinfo  {journal}
  {Nature}\ }\textbf {\bibinfo {volume} {412}},\ \bibinfo {pages} {295}
  (\bibinfo {year} {2001})}\BibitemShut {NoStop}%
\bibitem [{\citenamefont {Burt}\ \emph {et~al.}(1997)\citenamefont {Burt},
  \citenamefont {Ghrist}, \citenamefont {Myatt}, \citenamefont {Holland},
  \citenamefont {Cornell},\ and\ \citenamefont {Wieman}}]{Burt1997a}%
  \BibitemOpen
  \bibfield  {author} {\bibinfo {author} {\bibfnamefont {E.~A.}\ \bibnamefont
  {Burt}}, \bibinfo {author} {\bibfnamefont {R.~W.}\ \bibnamefont {Ghrist}},
  \bibinfo {author} {\bibfnamefont {C.~J.}\ \bibnamefont {Myatt}}, \bibinfo
  {author} {\bibfnamefont {M.~J.}\ \bibnamefont {Holland}}, \bibinfo {author}
  {\bibfnamefont {E.~A.}\ \bibnamefont {Cornell}}, \ and\ \bibinfo {author}
  {\bibfnamefont {C.~E.}\ \bibnamefont {Wieman}},\ }\href@noop {} {\bibfield
  {journal} {\bibinfo  {journal} {Phys. Rev. Lett}\ }\textbf {\bibinfo {volume}
  {79}},\ \bibinfo {pages} {337} (\bibinfo {year} {1997})}\BibitemShut
  {NoStop}%
\bibitem [{\citenamefont {Tolra}\ \emph {et~al.}(2004)\citenamefont {Tolra},
  \citenamefont {O'Hara}, \citenamefont {Huckans}, \citenamefont {Phillips},
  \citenamefont {Rolston},\ and\ \citenamefont {Porto}}]{Tolra2004a}%
  \BibitemOpen
  \bibfield  {author} {\bibinfo {author} {\bibfnamefont {B.~L.}\ \bibnamefont
  {Tolra}}, \bibinfo {author} {\bibfnamefont {K.~M.}\ \bibnamefont {O'Hara}},
  \bibinfo {author} {\bibfnamefont {J.~H.}\ \bibnamefont {Huckans}}, \bibinfo
  {author} {\bibfnamefont {W.~D.}\ \bibnamefont {Phillips}}, \bibinfo {author}
  {\bibfnamefont {S.~L.}\ \bibnamefont {Rolston}}, \ and\ \bibinfo {author}
  {\bibfnamefont {J.~V.}\ \bibnamefont {Porto}},\ }\href {\doibase
  10.1103/PhysRevLett.92.190401} {\bibfield  {journal} {\bibinfo  {journal}
  {Phys. Rev. Lett.}\ }\textbf {\bibinfo {volume} {92}},\ \bibinfo {pages}
  {190401} (\bibinfo {year} {2004})}\BibitemShut {NoStop}%
\bibitem [{\citenamefont {{Maier}}\ \emph {et~al.}(2015)\citenamefont
  {{Maier}}, \citenamefont {{Kadau}}, \citenamefont {{Schmitt}}, \citenamefont
  {{Wenzel}}, \citenamefont {{Ferrier-Barbut}}, \citenamefont {{Pfau}},
  \citenamefont {{Frisch}}, \citenamefont {{Baier}}, \citenamefont {{Aikawa}},
  \citenamefont {{Chomaz}}, \citenamefont {{Mark}}, \citenamefont {{Ferlaino}},
  \citenamefont {{Makrides}}, \citenamefont {{Tiesinga}}, \citenamefont
  {{Petrov}},\ and\ \citenamefont {{Kotochigova}}}]{Maier2015a}%
  \BibitemOpen
  \bibfield  {author} {\bibinfo {author} {\bibfnamefont {T.}~\bibnamefont
  {{Maier}}}, \bibinfo {author} {\bibfnamefont {H.}~\bibnamefont {{Kadau}}},
  \bibinfo {author} {\bibfnamefont {M.}~\bibnamefont {{Schmitt}}}, \bibinfo
  {author} {\bibfnamefont {M.}~\bibnamefont {{Wenzel}}}, \bibinfo {author}
  {\bibfnamefont {I.}~\bibnamefont {{Ferrier-Barbut}}}, \bibinfo {author}
  {\bibfnamefont {T.}~\bibnamefont {{Pfau}}}, \bibinfo {author} {\bibfnamefont
  {A.}~\bibnamefont {{Frisch}}}, \bibinfo {author} {\bibfnamefont
  {S.}~\bibnamefont {{Baier}}}, \bibinfo {author} {\bibfnamefont
  {K.}~\bibnamefont {{Aikawa}}}, \bibinfo {author} {\bibfnamefont
  {L.}~\bibnamefont {{Chomaz}}}, \bibinfo {author} {\bibfnamefont {M.~J.}\
  \bibnamefont {{Mark}}}, \bibinfo {author} {\bibfnamefont {F.}~\bibnamefont
  {{Ferlaino}}}, \bibinfo {author} {\bibfnamefont {C.}~\bibnamefont
  {{Makrides}}}, \bibinfo {author} {\bibfnamefont {E.}~\bibnamefont
  {{Tiesinga}}}, \bibinfo {author} {\bibfnamefont {A.}~\bibnamefont
  {{Petrov}}}, \ and\ \bibinfo {author} {\bibfnamefont {S.}~\bibnamefont
  {{Kotochigova}}},\ }\href@noop {} {\bibfield  {journal} {\bibinfo  {journal}
  {ArXiv e-prints}\ } (\bibinfo {year} {2015})},\ \Eprint
  {http://arxiv.org/abs/1506.05221} {arXiv:1506.05221 [cond-mat.quant-gas]}
  \BibitemShut {NoStop}%
\bibitem [{\citenamefont {Braaten}\ \emph {et~al.}(2002)\citenamefont
  {Braaten}, \citenamefont {Hammer},\ and\ \citenamefont
  {Mehen}}]{Braaten2002a}%
  \BibitemOpen
  \bibfield  {author} {\bibinfo {author} {\bibfnamefont {E.}~\bibnamefont
  {Braaten}}, \bibinfo {author} {\bibfnamefont {H.-W.}\ \bibnamefont {Hammer}},
  \ and\ \bibinfo {author} {\bibfnamefont {T.}~\bibnamefont {Mehen}},\ }\href
  {\doibase 10.1103/PhysRevLett.88.040401} {\bibfield  {journal} {\bibinfo
  {journal} {Phys. Rev. Lett.}\ }\textbf {\bibinfo {volume} {88}},\ \bibinfo
  {pages} {040401} (\bibinfo {year} {2002})}\BibitemShut {NoStop}%
\bibitem [{\citenamefont {Bulgac}(2002)}]{Bulgac2002a}%
  \BibitemOpen
  \bibfield  {author} {\bibinfo {author} {\bibfnamefont {A.}~\bibnamefont
  {Bulgac}},\ }\href {\doibase 10.1103/PhysRevLett.89.050402} {\bibfield
  {journal} {\bibinfo  {journal} {Phys. Rev. Lett.}\ }\textbf {\bibinfo
  {volume} {89}},\ \bibinfo {pages} {050402} (\bibinfo {year}
  {2002})}\BibitemShut {NoStop}%
\bibitem [{\citenamefont {Ronen}\ \emph {et~al.}(2006)\citenamefont {Ronen},
  \citenamefont {Bortolotti},\ and\ \citenamefont {Bohn}}]{Ronen2006a}%
  \BibitemOpen
  \bibfield  {author} {\bibinfo {author} {\bibfnamefont {S.}~\bibnamefont
  {Ronen}}, \bibinfo {author} {\bibfnamefont {D.~C.~E.}\ \bibnamefont
  {Bortolotti}}, \ and\ \bibinfo {author} {\bibfnamefont {J.~L.}\ \bibnamefont
  {Bohn}},\ }\href@noop {} {\bibfield  {journal} {\bibinfo  {journal} {Phys.
  Rev. A}\ }\textbf {\bibinfo {volume} {74}},\ \bibinfo {eid} {013623}
  (\bibinfo {year} {2006})}\BibitemShut {NoStop}%
\bibitem [{\citenamefont {Martin}\ and\ \citenamefont
  {Blakie}(2012)}]{Martin2012a}%
  \BibitemOpen
  \bibfield  {author} {\bibinfo {author} {\bibfnamefont {A.~D.}\ \bibnamefont
  {Martin}}\ and\ \bibinfo {author} {\bibfnamefont {P.~B.}\ \bibnamefont
  {Blakie}},\ }\href {\doibase 10.1103/PhysRevA.86.053623} {\bibfield
  {journal} {\bibinfo  {journal} {Phys. Rev. A}\ }\textbf {\bibinfo {volume}
  {86}},\ \bibinfo {pages} {053623} (\bibinfo {year} {2012})}\BibitemShut
  {NoStop}%
\bibitem [{\citenamefont {Blakie}\ \emph {et~al.}(2008)\citenamefont {Blakie},
  \citenamefont {Bradley}, \citenamefont {Davis}, \citenamefont {Ballagh},\
  and\ \citenamefont {Gardiner}}]{cfieldRev2008}%
  \BibitemOpen
  \bibfield  {author} {\bibinfo {author} {\bibfnamefont {P.~B.}\ \bibnamefont
  {Blakie}}, \bibinfo {author} {\bibfnamefont {A.~S.}\ \bibnamefont {Bradley}},
  \bibinfo {author} {\bibfnamefont {M.~J.}\ \bibnamefont {Davis}}, \bibinfo
  {author} {\bibfnamefont {R.~J.}\ \bibnamefont {Ballagh}}, \ and\ \bibinfo
  {author} {\bibfnamefont {C.~W.}\ \bibnamefont {Gardiner}},\ }\href@noop {}
  {\bibfield  {journal} {\bibinfo  {journal} {Adv. Phys.}\ }\textbf {\bibinfo
  {volume} {57}},\ \bibinfo {pages} {363} (\bibinfo {year} {2008})}\BibitemShut
  {NoStop}%
\bibitem [{\citenamefont {Tang}\ \emph {et~al.}(2015)\citenamefont {Tang},
  \citenamefont {Sykes}, \citenamefont {Burdick}, \citenamefont {Bohn},\ and\
  \citenamefont {Lev}}]{Tang2015a}%
  \BibitemOpen
  \bibfield  {author} {\bibinfo {author} {\bibfnamefont {Y.}~\bibnamefont
  {Tang}}, \bibinfo {author} {\bibfnamefont {A.}~\bibnamefont {Sykes}},
  \bibinfo {author} {\bibfnamefont {N.~Q.}\ \bibnamefont {Burdick}}, \bibinfo
  {author} {\bibfnamefont {J.~L.}\ \bibnamefont {Bohn}}, \ and\ \bibinfo
  {author} {\bibfnamefont {B.~L.}\ \bibnamefont {Lev}},\ }\href {\doibase
  10.1103/PhysRevA.92.022703} {\bibfield  {journal} {\bibinfo  {journal} {Phys.
  Rev. A}\ }\textbf {\bibinfo {volume} {92}},\ \bibinfo {pages} {022703}
  (\bibinfo {year} {2015})}\BibitemShut {NoStop}%
\bibitem [{Note1()}]{Note1}%
  \BibitemOpen
  \bibinfo {note} {For $a_f\gtrsim 92\protect \tmspace +\thinmuskip
  {.1667em}a_0$ we do not find any droplets forming within $15\protect \tmspace
  +\thinmuskip {.1667em}m$s, and on longer timescales much few droplets form
  than observed in experiment.}\BibitemShut {Stop}%
\bibitem [{\citenamefont {Santos}\ \emph {et~al.}(2003)\citenamefont {Santos},
  \citenamefont {Shlyapnikov},\ and\ \citenamefont {Lewenstein}}]{Santos2003a}%
  \BibitemOpen
  \bibfield  {author} {\bibinfo {author} {\bibfnamefont {L.}~\bibnamefont
  {Santos}}, \bibinfo {author} {\bibfnamefont {G.~V.}\ \bibnamefont
  {Shlyapnikov}}, \ and\ \bibinfo {author} {\bibfnamefont {M.}~\bibnamefont
  {Lewenstein}},\ }\href {\doibase 10.1103/PhysRevLett.90.250403} {\bibfield
  {journal} {\bibinfo  {journal} {Phys. Rev. Lett.}\ }\textbf {\bibinfo
  {volume} {90}},\ \bibinfo {pages} {250403} (\bibinfo {year}
  {2003})}\BibitemShut {NoStop}%
\bibitem [{\citenamefont {Ronen}\ \emph {et~al.}(2007)\citenamefont {Ronen},
  \citenamefont {Bortolotti},\ and\ \citenamefont {Bohn}}]{Ronen2007a}%
  \BibitemOpen
  \bibfield  {author} {\bibinfo {author} {\bibfnamefont {S.}~\bibnamefont
  {Ronen}}, \bibinfo {author} {\bibfnamefont {D.~C.~E.}\ \bibnamefont
  {Bortolotti}}, \ and\ \bibinfo {author} {\bibfnamefont {J.~L.}\ \bibnamefont
  {Bohn}},\ }\href {\doibase 10.1103/PhysRevLett.98.030406} {\bibfield
  {journal} {\bibinfo  {journal} {Phys. Rev. Lett.}\ }\textbf {\bibinfo
  {volume} {98}},\ \bibinfo {eid} {030406} (\bibinfo {year}
  {2007})}\BibitemShut {NoStop}%
\bibitem [{\citenamefont {Blakie}\ \emph {et~al.}(2012)\citenamefont {Blakie},
  \citenamefont {Baillie},\ and\ \citenamefont {Bisset}}]{Blakie2012a}%
  \BibitemOpen
  \bibfield  {author} {\bibinfo {author} {\bibfnamefont {P.~B.}\ \bibnamefont
  {Blakie}}, \bibinfo {author} {\bibfnamefont {D.}~\bibnamefont {Baillie}}, \
  and\ \bibinfo {author} {\bibfnamefont {R.~N.}\ \bibnamefont {Bisset}},\
  }\href {\doibase 10.1103/PhysRevA.86.021604} {\bibfield  {journal} {\bibinfo
  {journal} {Phys. Rev. A}\ }\textbf {\bibinfo {volume} {86}},\ \bibinfo
  {pages} {021604} (\bibinfo {year} {2012})}\BibitemShut {NoStop}%
\bibitem [{\citenamefont {Kim}\ and\ \citenamefont {Chan}(2004)}]{Kim2004a}%
  \BibitemOpen
  \bibfield  {author} {\bibinfo {author} {\bibfnamefont {E.}~\bibnamefont
  {Kim}}\ and\ \bibinfo {author} {\bibfnamefont {M.~H.~W.}\ \bibnamefont
  {Chan}},\ }\href {http://dx.doi.org/10.1038/nature02220} {\bibfield
  {journal} {\bibinfo  {journal} {Nature}\ }\textbf {\bibinfo {volume} {427}},\
  \bibinfo {pages} {225} (\bibinfo {year} {2004})}\BibitemShut {NoStop}%
\bibitem [{\citenamefont {Kim}\ and\ \citenamefont {Chan}(2012)}]{Kim2012a}%
  \BibitemOpen
  \bibfield  {author} {\bibinfo {author} {\bibfnamefont {D.~Y.}\ \bibnamefont
  {Kim}}\ and\ \bibinfo {author} {\bibfnamefont {M.~H.~W.}\ \bibnamefont
  {Chan}},\ }\href {\doibase 10.1103/PhysRevLett.109.155301} {\bibfield
  {journal} {\bibinfo  {journal} {Phys. Rev. Lett.}\ }\textbf {\bibinfo
  {volume} {109}},\ \bibinfo {pages} {155301} (\bibinfo {year}
  {2012})}\BibitemShut {NoStop}%
\bibitem [{\citenamefont {Boninsegni}\ and\ \citenamefont
  {Prokof'ev}(2012)}]{Boninsegni2012a}%
  \BibitemOpen
  \bibfield  {author} {\bibinfo {author} {\bibfnamefont {M.}~\bibnamefont
  {Boninsegni}}\ and\ \bibinfo {author} {\bibfnamefont {N.~V.}\ \bibnamefont
  {Prokof'ev}},\ }\href {\doibase 10.1103/RevModPhys.84.759} {\bibfield
  {journal} {\bibinfo  {journal} {Rev. Mod. Phys.}\ }\textbf {\bibinfo {volume}
  {84}},\ \bibinfo {pages} {759} (\bibinfo {year} {2012})}\BibitemShut
  {NoStop}%
\bibitem [{\citenamefont {{Xi}}\ and\ \citenamefont {{Saito}}(2015)}]{Xi2015a}%
  \BibitemOpen
  \bibfield  {author} {\bibinfo {author} {\bibfnamefont {K.-T.}\ \bibnamefont
  {{Xi}}}\ and\ \bibinfo {author} {\bibfnamefont {H.}~\bibnamefont {{Saito}}},\
  }\href@noop {} {\bibfield  {journal} {\bibinfo  {journal} {ArXiv e-prints}\ }
  (\bibinfo {year} {2015})},\ \Eprint {http://arxiv.org/abs/1510.07842}
  {arXiv:1510.07842 [cond-mat.quant-gas]} \BibitemShut {NoStop}%
\end{thebibliography}
%

\end{document}